\documentclass[
superscriptaddress,
amsmath,amssymb,
aps,
prb,
twocolumn,
]{revtex4-2}

\usepackage{graphicx}
\usepackage{dcolumn}
\usepackage{bm}
\usepackage{bbold} 
\usepackage[breaklinks,colorlinks = true,linkcolor = magenta,urlcolor=magenta,citecolor=red]{hyperref}
\usepackage{url}
\usepackage{tikz}
\usepackage{diagbox}
\usepackage{multirow}
\usepackage{hhline}
\usepackage{array}
\usepackage{booktabs}
\usepackage{ctable}
\usepackage{upgreek}
\usepackage{epsfig,psfrag,subfigure,amsopn}
\usepackage{mathrsfs}
\usepackage{amssymb}
\usepackage{amsbsy}
\usepackage{color}
\usepackage{cancel}
\usepackage{pifont}
\usepackage{marginnote}
\usepackage{float}
\usepackage{physics}
\usepackage[final]{showlabels} 
\usepackage{soul}

\newcommand{\beq}{\begin{equation}}
\newcommand{\eeq}{\end{equation}}
\newcommand{\beqar}{\begin{eqnarray}}
\newcommand{\eeqar}{\end{eqnarray}}

\newcommand{\veryshortarrow}[1][3pt]{\mathrel{%
   \hbox{\rule[\dimexpr\fontdimen22\textfont2-.2pt\relax]{#1}{.4pt}}%
   \mkern-4mu\hbox{\usefont{U}{lasy}{m}{n}\symbol{41}}}}
\newcommand{\Co}{{c}}
\newcommand{\Ho}{{h}}

\newcommand{\pii}{\Pi^{i_\Co i_\Ho}}

\begin{document}
\relax



\title{Probing quantum anomalous heat flow using mid-circuit measurements}

\author{Aabhaas Vineet Mallik}
\email{aabhaas.iiser@gmail.com}
\affiliation{Department of Physics, Bar-Ilan University, 52900 Ramat Gan, Israel}
\affiliation{Department of Physics, BITS Pilani K K Birla Goa Campus, Zuarinagar, Goa, India - 403726}

\author{Loris Maria Cangemi}
\altaffiliation{Present address: Dept. of Electrical Engineering and Information Technology,
Università degli Studi di Napoli Federico II, via Claudio 21, Napoli, 80125, Italy.}
\affiliation{Department of Chemistry, Bar-Ilan University, Ramat-Gan, 52900 Israel}
\affiliation{Institute of Nanotechnology and Advanced Materials,  Bar-Ilan University, 52900 Ramat Gan, Israel}
\affiliation{Center for Quantum Entanglement Science and Technology, Bar-Ilan University, 52900 Ramat Gan, Israel}

\author{Amikam Levy}
\affiliation{Department of Chemistry, Bar-Ilan University, Ramat-Gan, 52900 Israel}
\affiliation{Institute of Nanotechnology and Advanced Materials,  Bar-Ilan University, 52900 Ramat Gan, Israel}
\affiliation{Center for Quantum Entanglement Science and Technology, Bar-Ilan University, 52900 Ramat Gan, Israel}

\author{Emanuele G. Dalla Torre}
\affiliation{Department of Physics, Bar-Ilan University, 52900 Ramat Gan, Israel}
\affiliation{Center for Quantum Entanglement Science and Technology, Bar-Ilan University, 52900 Ramat Gan, Israel}



\date{\today}

\begin{abstract}
Gate-based quantum computers are an innovative tool for experimentally studying the core principles of quantum mechanics. This work presents the first observation of quantum anomalous heat flow between two qubits and investigates the role of mid-circuit measurements in this context. Using mid-circuit measurements, we designed quantum circuits that violate the semi-classical heat flow bound, witnessing negativities in the underlying Kirkwood-Dirac quasiprobability distribution, which indicates the presence of quantum correlations between the subsystems. Mid-circuit measurements, crucial for probing qubits during the experiment, enabled these observations but also introduced disturbances, such as energy leakage, leading to deviations from theoretical predictions. We modeled these noise effects, providing insight into the limitations of current mid-circuit measurement techniques. 

\end{abstract}

\maketitle


\section{Introduction}
State-of-the-art quantum computers provide access to processors with a relatively large number of qubits. The individual qubits, however, suffer from significant decoherence effects which make the realization
of specific many-body quantum states (algorithms) very challenging. 
Correcting these errors requires 
many physical qubits for each logical one and
is not practically relevant for computational tasks on noisy intermediate scale quantum (NISQ) devices. For these systems, one usually tries to achieve an easier task, namely to mitigate the effects of noise, rather than correcting it \cite{CaiQEMreview_2023}. Mitigation techniques have found important applications in the simulation of 
interesting quantum many-body effects, such as, time crystals \cite{mi2022time}, relativistic wormholes \cite{jafferis2022traversable}, and the dynamics of spin models \cite{2023_Nat_Kim}, and many more \cite{alexeev2021quantum}.

Recently, a new versatile tool of \emph{mid-circuit measurement}, the ability to measure a set of selected qubits during the evolution of a quantum circuit and then to use the output of these measurements for further classical and quantum computations, has been introduced \cite{IBMFalcon}. Stability of quantum computers to mid-circuit measurements can potentially be useful in the implementation of quantum error correction protocols \cite{Rudinger_2022,Botelho_2022} and hence, is an important milestone in the progress of quantum computation technologies. In this paper, we demonstrate that quantum computers stable to mid-circuit measurements also become amenable to some exquisite quantum thermodynamics experiments.

Problems in quantum thermodynamics are of great contemporary interest both from the point of view of fundamental physics as well as applications in engineering quantum devices \cite{kosloff2014quantum,goold2016role,vinjanampathy2016quantum,levy2018quantum,deffner2019quantum,2024_arXiv_Cleri,CANGEMI20241,2023_JMRO_Vieira}. As a result, a wide variety of experimental platforms have been used to explore different aspects of these problems \cite{myers2022quantum, Arrachea_2023,CANGEMI20241,2023_JMRO_Vieira}. The use of NISQ computers to study quantum thermodynamics is promising \cite{2021_PRXQ_Andrea, oftelie2022computing} and remains largely unexplored. In this study, we present findings from a series of quantum thermodynamics experiments conducted on NISQ computers, which enable us to investigate the impact of initial quantum correlations on the heat flow between two parts of a quantum system.

In particular, we consider the heat exchange between two subsystems of a quantum device whose initial state is chosen to generalize heat flow experiments at the macroscopic scale. This is achieved by working with a particular class of initial states whose associated density matrix $\rho_{ch}$ has thermal marginals with different temperatures for each of the two subsystems. Naturally, the subsystem with the lower temperature can be referred to as cold ($c$) and the other one as hot ($h$). Note that $\rho_{ch}$ can account for both classical and quantum correlations between the subsystems. The evolution of the quantum device is generically governed by the Hamiltonian
\begin{equation}
    H_\theta = H_c + H_h + H_{ex}(\theta).
    \label{eq:hamiltonian}
\end{equation}
Here, $H_{c(h)}$ governs the free evolution of the cold (hot) subsystem,  while $H_{ex}(\theta)$ describes the interactions between $c$ and $h$ facilitating the exchange of heat and is controlled by a set of parameters $\theta$. We further constrain $H_{ex}(\theta)$ such that the heat exchanged between $c$ and $h$ in a unit time interval conserves the combined energy of the two subsystems. That is, the unitary evolution of the quantum device over a unit time interval, $U_\theta = \exp(-i H_\theta )$, satisfies
\begin{equation}
    [U_\theta,H_c+H_h] = 0 \ .
    \label{eq:energy_cond}
\end{equation}
Eq.~(\ref{eq:energy_cond}) ensures that the heat exchange between $c$ and $h$ is well defined.


In this study, we designed a series of experiments on IBM quantum computers to observe quantum anomalous heat flow between two qubits. This phenomenon is witnessed by violating a semi-classical bound described as a heat flow inequality, as detailed in~\cite{2020_PRXQ_Levy}. To determine this bound, we employed mid-circuit measurements, which facilitate the assessment of heat flow through a two-point measurement scheme.
In addition to marking the first observation of quantum anomalous heat flow on quantum computers, our findings reveal that the discrepancies between experimental results and theoretical predictions are primarily due to mid-circuit measurements. These measurements introduce disturbances that result in energy leakage and biased outcomes. We modeled this noise by adjusting the population distribution by a fixed amount, improving the agreement between our theoretical framework and the experimental data.

The structure of this manuscript is as follows. In Sec.~\ref{sec:heat flow}, we discuss the connection between anomalous heat flow and quasiprobability distributions and outline the heat flow inequality. We then introduce the experiment conducted on the IBM quantum computers and provide details on the quantum circuits used to estimate the semi-classical bound on the heat flow. Sec.~\ref{sec:results} presents the experimental results from these quantum circuits, including the detection of anomalous heat flow and an analysis of the impact of mid-circuit measurements. Lastly, in Sec.~\ref{sec: summary}, we provide a brief summary and discussion of the key findings.

\section{Heat flow between two qubits}
\label{sec:heat flow}

\begin{figure}
    \centering
\includegraphics[width=1\columnwidth,trim={3cm 0 0 0},clip]{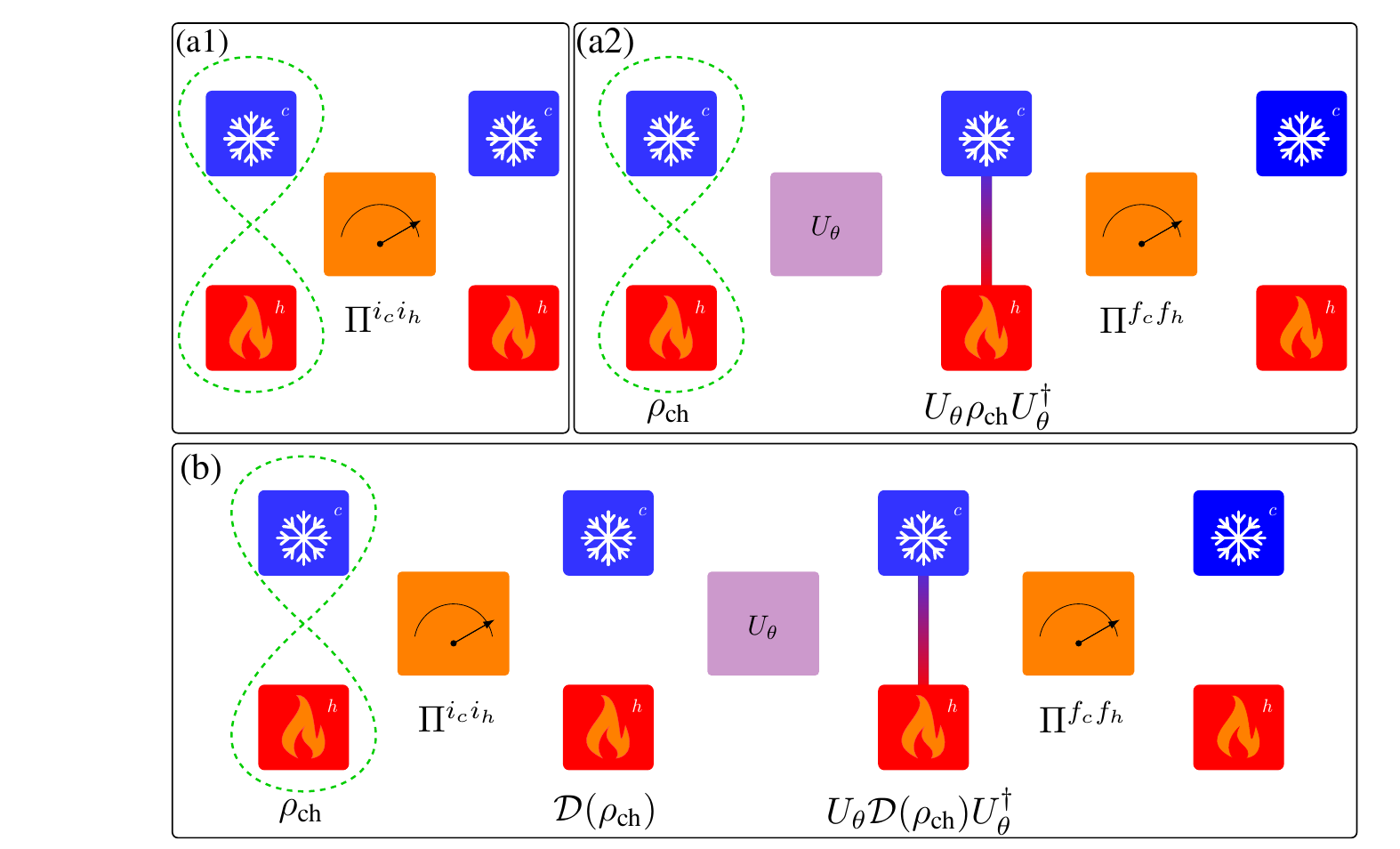}
    \caption{\emph{Schematics of the heat flows measurement process:} The red and the blue blocks together represent the quantum device with the blue (red) block representing the subsystem which started with a lower (higher) effective temperature. Panels (a1) and (a2) describe the processes for evaluating $Q_{\rm q}$ defined in Eq.~(\ref{eq:formula_Qqc}). Panel (b) describes the process for evaluating $Q_{\rm sc}$ defined in Eq.~(\ref{eq:Qsc}) using a two-point measurement (TPM) scheme.}
    \label{fig:scheme}
\end{figure}

\subsection{Heat flow and quasiprobability distributions}
When two local thermal subsystems are initially uncorrelated, the heat flowing between them is governed by Jarzynski's heat exchange fluctuation theorem~\cite{jarzynski2004classical}, which mandates that on average, heat will flow from the hot system to the cold one. Interestingly, in the presence of initial correlations, classical or quantum, heat can flow backward from cold to hot, on average~\cite{lloyd1989use,jennings10}. If the initial correlations are purely classical \footnote{The heat exchange process can still be quantum as quantum correlation can be created during the unitary evolution.}, the process is governed by a classical joint probability distribution~\cite{jennings12},
whereas in the presence of initial quantum correlations, a quasiprobability distribution is required to describe the heat exchange statistics~\cite{2020_PRXQ_Levy,lostaglio2023kirkwood}.
This distinction originates from the incompatibility of observables in quantum mechanics that prevents representing a quantum state or a process using a joint probability distribution. 

Although no unique quasiprobability distribution characterizes processes in the quantum realm,
the Kirkwood-Dirac quasiprobability (KDQ)~\cite{ kirkwood1933quantum,dirac1945analogy} was suggested as a natural choice for the statistics of the heat exchange process and the generalization of Jarzynski's heat exchange fluctuation theorem to the full quantum regime~\cite{2020_PRXQ_Levy}. 
The KDQ can take negative and non-real values and has seen a resurgence of interest due to its implications to witnessing non-classicality and its applications across various fields of physics, including foundations of quantum mechanics~\cite{SpekkensPRL2008,HofmannPRL2012,PuseyPRL2014,lupu2021negative,Dressel2014colloquium,KunjwalPRA2019}, quantum chaos and many-body physics~\cite{yunger2018quasiprobability,dressel2018strengthening,mohseninia2019strongly,alonso2019out}, quantum thermodynamics~\cite{allahverdyan2014nonequilibrium,2020_PRXQ_Levy,hernandez2024projective,PeiPRE2023,GherardiniTutorial,hernandez2024interferometry} 
and more~\cite{DeBievrePRL2021,BudiyonoPRAquantifying,lostaglio2023kirkwood,wagner2024quantum,arvidsson2024properties}. 

In the heat exchange process between two correlated states, the heat flowing out of the cold (hot) subsystem $c(h)$,
\begin{equation}
Q_{\rm q}^{c(h)} = {\rm Tr} \left[ \rho_{ch} H_{c(h)} - U_\theta\rho_{ch} U^\dagger _\theta H_{c(h)}\right],
\label{eq:formula_Qqc}
\end{equation}
can be expressed in terms of the real part of the KDQ,
\begin{equation}
    Q_{\rm q}^{c(h)}=\sum_{i_c,i_h,f_c,f_h}p_{i_c,i_h\veryshortarrow f_c,f_h}(E_{i_{c(h)}}-E_{f_{c(h)}}).
\end{equation}
Here $p_{i_c,i_h\veryshortarrow f_c,f_h}={\rm Re Tr}[\tilde{\Pi}^{f_{c}f_{h}} \pii \rho_{ch}]$ is the real part of KDQ (also known as the Margenau-Hill distribution~\cite{margenau1961correlation}), and $\pii\equiv \Pi^{i_c}\otimes \Pi^{i_h}$ where $\Pi^{i_x}=\ket{E_{i_x}}\bra{E_{i_x}}$ are energy-projection operators such that $H_x=\sum_{i_x} E_{i_x} \Pi^{i_x} $  for $x=c,h$. The subscripts $i$ and $f$ refer to the initial and final energies and $\tilde{\Pi}=U_{\theta}^{\dagger}\Pi U_{\theta}$ is the propagated projection operator.
Since the heat that flows out of the cold subsystems equals the heat that flows into the hot system, {\it i.e.} $Q_{\rm q}^c=-Q_{\rm q}^h$, we will use the single notation and convention $Q_{\rm q}>0$ for heat flowing from the cold subsystem to the hot one.  
To evaluate $Q_{\rm q}$, two separate measurements are performed, evaluating the first and second term inside the trace in Eq.~(\ref{eq:formula_Qqc}) as schematically shown in Fig.~\ref{fig:scheme} panels (a1) and (a2) respectively.
In both cases, the initial state is the same correlated state with thermal marginals. The measurement in panel (a1) provides the initial local energy of the subsystems. In contrast, panel (a2) provides the local energy of the subsystems after evolving the initial state and letting the subsystems exchange heat.

Nonclassicality in the heat exchange process can be linked to instances where the quasiprobability turns negative. 
Ref.~\cite{2020_PRXQ_Levy} provides an explicit inequality whose violation is a witness of quantumness arising from the negativity of the quasiprobability, and for a two-qubit system it reads
\begin{equation}
    I \equiv |Q_{\rm q}| - \frac{2+e^{\beta_c \Delta}+e^{\beta_h \Delta}}{e^{\beta_c \Delta}-e^{\beta_h \Delta}} |Q_{\rm sc}| \leq 0 \ .
    \label{eq:2quibit_ineq}
\end{equation}
Here $\beta_c$ and $\beta_h$ are the local inverse temperatures of the cold and hot qubits, and $\Delta$ is the energy gap of the qubits (we take $\hbar=k_B=1$). 
The semi-classical heat flow $Q_{\rm sc}$ is
the heat exchanged between the two qubits in a two-point measurement (TPM) scheme~\cite{campisi2011colloquium} featured in  Fig.\ref{fig:scheme}(b). In this scheme, 
a first projective (mid-circuit)  measurement of the energy is carried out, then, the post-measurement state undergoes a unitary evolution in which heat is exchanged between the two subsystems, and finally, a second energy projective measurement is performed. 
This scheme results in, 
\begin{align}
    Q_{\rm sc}^{c(h)}=\Tr[\rho_{ch}H_{c(h)} - U_{\theta}\mathcal{D}\rho_{ch}U_{\theta}^{\dagger}H_{c(h)}].
    \label{eq:Qsc}
\end{align}
Here, $\mathcal{D}\rho_{ch}$ is the dephased initial state in the energy basis, describing the average state of the system following the first projective measurement that eliminates all initial quantum correlations and phases in the energy basis. To witness nonclassicality we will observe the violation of the inequality in Eq.~(\ref{eq:2quibit_ineq}) by measuring $Q_{\rm q}$ and $Q_{sc}$ separately. When this inequality is violated, the quantum heat flow $Q_{\rm q}$ is referred to as the \emph{quantum anomalous heat flow}. 

\subsection{Probing heat flow using quantum circuits}
\begin{figure}
    \centering
\includegraphics[width=0.8\columnwidth]{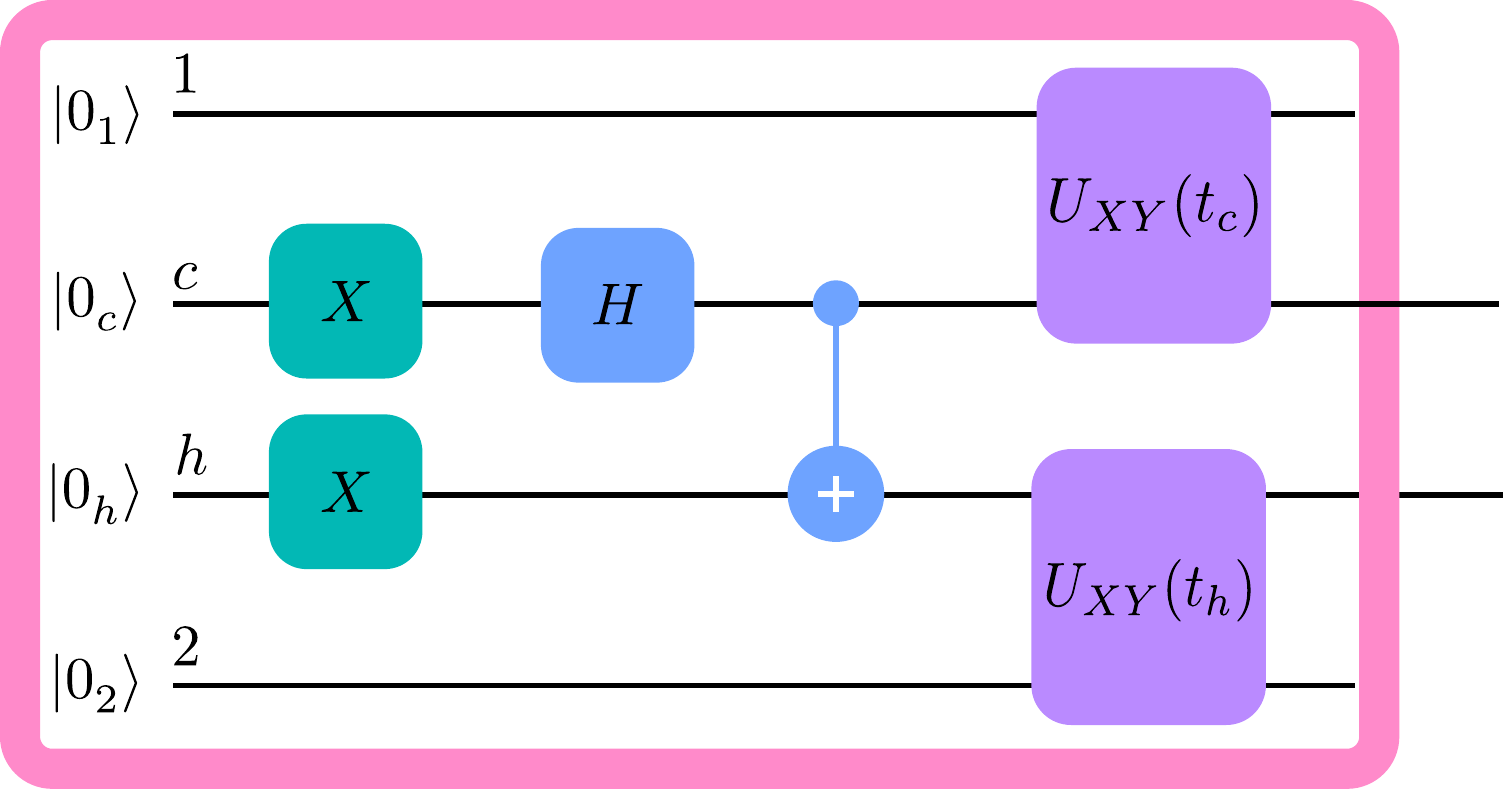}
    \caption{{\it Preparation of the initial state}: Qubits $c$ and $h$ together form the quantum device for the heat flow experiments, while qubits $1$ and $2$ are auxiliary qubits. The direction of forward evolution is from left to right. All the qubits are initiated in their ground states $|0_i\rangle$ where $i$ is the index of the qubit. See text for a detailed discussion of the gates used.}
    \label{fig:init_state}
\end{figure}

We consider the heat exchange between two qubits in resonance, such that the energy gap $\Delta$ between the ground states $|0_c\rangle$ and $|0_h\rangle$ and the excited states $|1_c\rangle$ and $|1_h\rangle$ of the cold and the hot qubits, respectively, is the same. 
This is a necessary condition for no-work injection during the heat exchange process. The first challenge that we address is how to initialize the two qubits in local thermal states while keeping quantum correlations between them. We achieve this goal in a controlled manner with the help of two auxiliary qubits and two unitary gates, parameterized by the parameters $t_c$ and $t_h$. The protocol that we use is shown in Fig.~\ref{fig:init_state} and consists of the following steps.

As a first step, after initializing all the qubits in their ground states, we prepare the qubits $c$ and $h$ in the singlet state $(|0_c1_h \rangle - |1_c 0_h\rangle)/\sqrt{2}$ using three one-qubit gates, (a bit-flip ($X$) on $c$ and $h$, followed by a Hadamard ($H$) on $c$) and a controlled-NOT gate (where $h$ is flipped depending on the state of $c$). At this stage, the marginal state of each qubit is maximally mixed and corresponds to an ``infinite temperature'' ensemble. 

\begin{figure}
    \centering
\includegraphics[width=0.8\columnwidth]{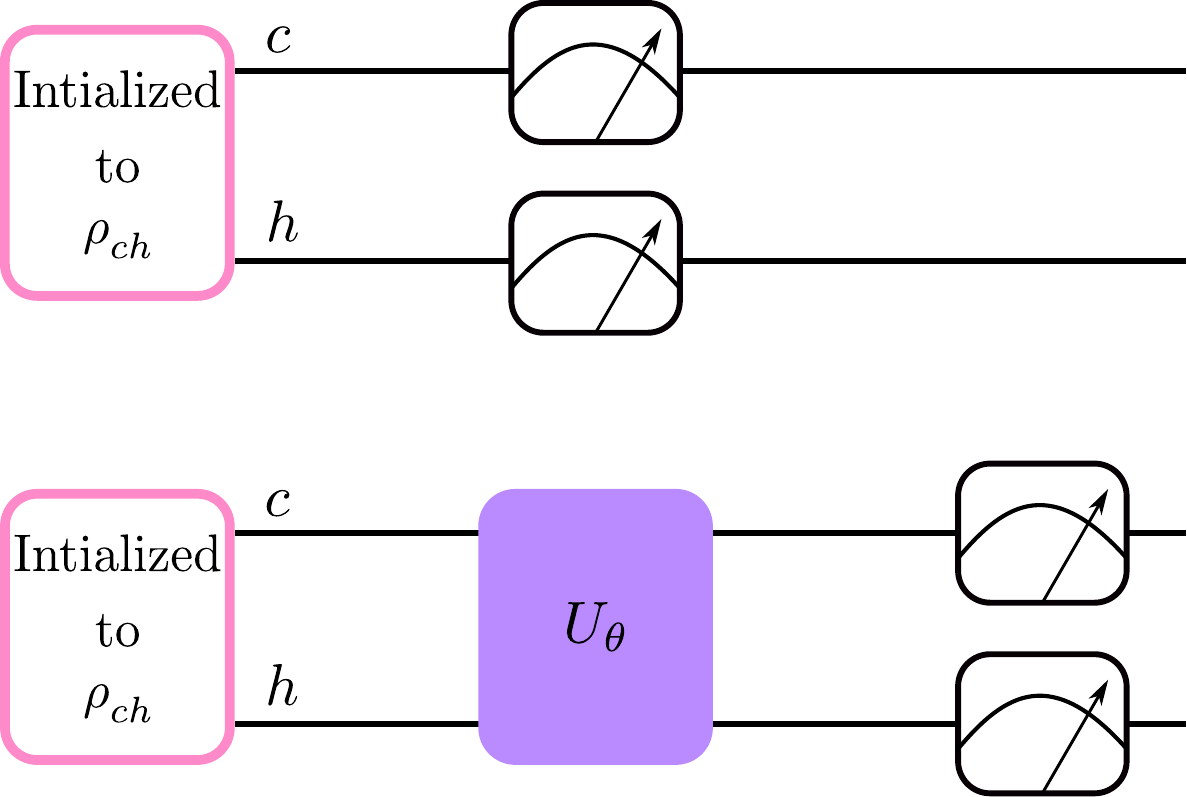}
    \caption{{\it Quantum circuit to measure $Q_{\rm q}$ for our two qubit system}: As described in Fig.~\ref{fig:scheme}(a), measurements on two independent circuits are required to estimate $Q_{\rm q}$. The first circuit (top) consists of the preparation circuit shown in Fig.~\ref{fig:init_state} which initializes the two qubits in the desired state $\rho_{ch}$, followed by projective measurements of the c and h qubits in the canonical basis. Repeating these measurements on identically prepared qubits provides $\langle E_{i_c} \rangle$ and $\langle E_{i_h} \rangle$. The second circuit (bottom) includes the unitary operation $U_\theta$ and provides $\langle E_{f_c} \rangle$ and $\langle E_{f_h} \rangle$. From these quantities we compute $Q_{\rm q} = \langle E_{i_c} \rangle -  \langle E_{f_c}\rangle = \langle E_{f_h} \rangle - \langle E_{i_h}\rangle$.}
    \label{fig:qct_circuit}
\end{figure}

\begin{figure}
    \centering
\includegraphics[width=0.8\columnwidth]{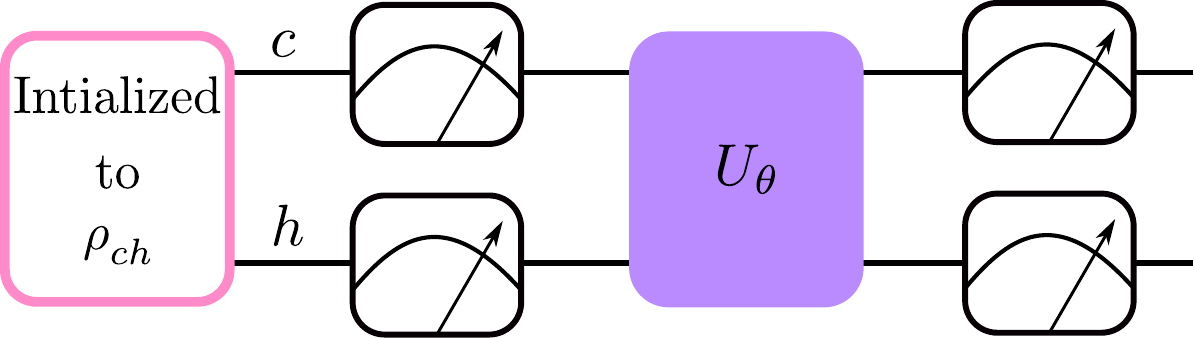}
    \caption{{\it Quantum circuit to measure $Q_{\rm sc}$ for our two qubit system}: As described in Fig.~\ref{fig:scheme}(b), measurements on a single circuit is sufficient to estimate $Q_{\rm sc}$. The first block initializes the qubits in the desired state $\rho_{ch}$ (see Fig.~\ref{fig:init_state}). The first set of measurements provides $E_{i_c}$ and $E_{i_h}$ while the final set of measurements provide $\tilde{E}_{f_c}$ and $\tilde{E}_{f_h}$. Implementing this circuit several times allows one to estimate $Q_{\rm sc} = \langle E_{i_c} - \tilde{E}_{f_c}\rangle = \langle \tilde{E}_{f_h} - E_{i_h}\rangle$.}
    \label{fig:qct_tpm_circuit}
\end{figure}

The next step allows us to prepare the qubits in thermal marginals with arbitrary temperatures $\beta^{-1}_c$ and $\beta^{-1}_h$. This is achieved by entangling $c$ with qubit $1$ and $h$ with qubit $2$ using a two-qubit gate  $U_{XY}(t)$, which performs a parameterized swap operation. With the two-qubit basis, $|00\rangle$, $|01\rangle$, $|10\rangle$ and $|11\rangle$ represented by the standard basis $\mathbf{e}_1$, $\mathbf{e}_2$, $\mathbf{e}_3$ and $\mathbf{e}_4$, respectively,
\begin{equation}
    U_{XY}(t) = 
    \begin{pmatrix}
    1 & 0 & 0 & 0 \\
    0 & \cos (t/2) & i \sin (t/2) & 0 \\
    0 & i\sin (t/2) & \cos (t/2) & 0 \\
    0 & 0 & 0 & 1
    \end{pmatrix} \ .
\end{equation}
As $t$ is varied from zero to $\pi$, $U_{XY}(t)$ varies from being identity to a complete swap.
If $U_{XY}(t_c)$ is used to swap $1$ and $c$, while $U_{XY}(t_h)$ is used to swap $2$ and $h$ then the qubits $c$ and $h$ end up in a state described by a density matrix of the form
\begin{equation}
    \rho_{ch} = \begin{pmatrix}
                P_{00} & 0 & 0 & 0 \\
                0 & \frac{1}{z_c} - P_{00} & \eta & 0 \\
                0 & \eta & \frac{1}{z_h} - P_{00} & 0 \\
                0 & 0 & 0 & 1 - \frac{1}{z_c} - \frac{1}{z_h} + P_{00} \end{pmatrix}
    \label{eq:init_state}
\end{equation}
where
\begin{gather}
    P_{00} = 1 - \frac{1}{2}\left( \cos ^2\left(\frac{t_c}{2}\right) + \cos ^2\left(\frac{t_h}{2}\right) \right) \label{eq:P00} \\
    \eta = -\frac{1}{2}\cos\left(\frac{t_c}{2}\right)\cos\left(\frac{t_h}{2}\right) \label{eq:eta} \\
    \frac{1}{z_{c(h)}} = 1 - \frac{1}{2} \cos^2\left(\frac{t_{c(h)}}{2}\right)\ \label{eq:partition_func}.
\end{gather}
Clearly, the reduced density matrices,
\begin{equation}
    \rho_{c(h)} = {\rm Tr}_{h(c)} \rho_{ch} = \left[\begin{matrix}1/z_{c(h)} & 0\\0 & 1 - 1/z_{c(h)}\end{matrix}\right]
    \label{eq:rho_c}
\end{equation}
look thermal. Specifically, identifying $z_{c(h)}$ with $1+\exp(-\beta_{c(h)} \Delta)$, implies that the effective inverse temperature of the qubit $c(h)$,
\begin{equation}
    \beta_{c(h)} = \frac{1}{\Delta}\ln \left( 1+2\tan^2\left( \frac{t_{c(h)}}{2} \right) \right) \ ,
    \label{eq:eff_temp}
\end{equation}
can be set to the desired value by choosing $t_{c(h)}$ appropriately.

The exchange of energy between $c$ and $h$ is facilitated by applying a unitary operator $U_\theta$ that
allows these qubits to interact, while keeping the sum of their energies constant (see Eq.~(\ref{eq:energy_cond})). As shown in Ref.~\cite{2020_PRXQ_Levy}, this two-qubit gate can be parameterized by a single parameter $\theta$
\begin{equation}
U_\theta = \begin{pmatrix}
1 & 0 & 0 & 0 \\
0 & \cos{\theta} & -\sin{\theta} & 0 \\
0 & \sin{\theta} & \cos{\theta} & 0 \\
0 & 0 & 0 & 1
\end{pmatrix}.
\label{eq:U_theta}
\end{equation}
The heat exchanged between the hot and cold subsystems is fully determined by the parameters $t_c$, $t_h$, and $\theta$.

\begin{figure}
    \centering
\includegraphics[width=1.0\columnwidth]{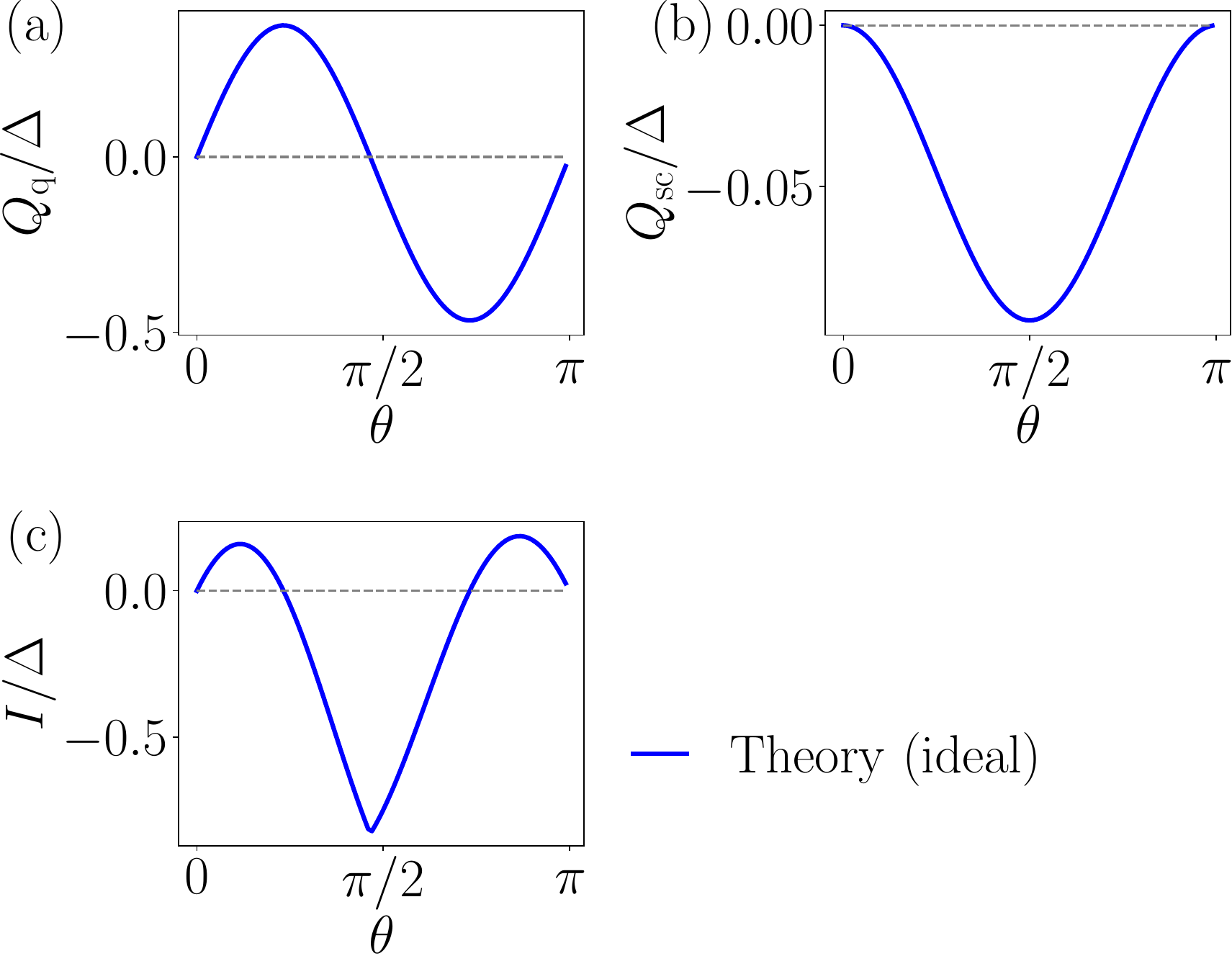}
    \caption{{\it Theoretical prediction}: For $\beta_c \Delta =0.51$ and $\beta_h \Delta =0.13$, (a) and (b) show the theoretical values of $Q_{\rm q}/\Delta$ and $Q_{\rm sc}/\Delta$, respectively, as a function of $\theta$ which parameterizes the energy preserving unitary evolution $U_\theta$ (for more details see text). (c) shows the violation of the inequality Eq.~(\ref{eq:2quibit_ineq}) for values of $\theta$ close to zero and $\pi$. The quantum heat flow $Q_{\rm q}$ when Eq.~(\ref{eq:2quibit_ineq}) is violated is referred to as the \emph{quantum anomalous heat flow}.}
    \label{fig:th}
\end{figure}

The quantum and semi-classical heat flows can be computed analytically \cite{2020_PRXQ_Levy} and are respectively equal to
\begin{align}
    \frac{Q_{\rm q}}{\Delta} &= \frac{\cos^{2}{\left(\frac{t_{c}}{2} \right)}}{2} - \frac{\left(\cos{\left(\frac{t_{c}}{2} \right)} \cos{\left(\theta \right)} - \sin{\left(\theta \right)} \cos{\left(\frac{t_{h}}{2} \right)} \right)^{2}}{2},\nonumber\\
    \frac{Q_{\rm sc}}{\Delta} &= \frac{1}{4} \sin^2{\left(\theta \right)} \left( \cos\left(t_{c}\right) - \cos\left(t_{h}\right) \right)
    \label{eq:Qtpm_t1t2theta} \ .
\end{align}
These quantities can be measured in a quantum computer using the quantum circuits shown in Figs.~\ref{fig:qct_circuit} and \ref{fig:qct_tpm_circuit}. The former uses two distinct circuits to measure the quantum heat flow $Q_{\rm q}$, while the latter uses a single circuit with mid-circuit measurements to probe the semi-classical heat flow $Q_{\rm sc}$. For all the results presented in this paper, the two-qubit state is initialized with the choice $t_c=\pi/3$ and $t_h=\pi/6$, that is, $\beta_c \Delta =0.51$ and $ \beta_h \Delta=0.13$ (see Eq.~(\ref{eq:eff_temp})).
For these selected parameters, we plot $Q_{\rm q}/\Delta$ and $Q_{\rm sc}/\Delta$ as functions of $\theta$ in Fig.~\ref{fig:th}(a) and (b), respectively. Additionally, in Fig.~\ref{fig:th}(c), we plot the corresponding quantum violation $I$ for various values of the evolution parameter~$\theta$.

\subsection{IBM quantum processing units}
The proposed implementation requires mid-circuit measurements, which can introduce noise and potentially compromise the subsequent evolution of the quantum system and the accuracy of the theoretically predicted quantum violation. It is, of course, desirable to obtain a qualitative and, if possible, quantitative understanding of these effects on the measurements of $Q_{\rm q}$ and $Q_{\rm sc}$. To this end, it is crucial to select a platform that offers access to multiple realizations of NISQ computers varying in their stability to mid-circuit measurements.

IBM Quantum currently serves as a leading platform providing such access, and all subsequent results presented are derived from computations conducted on various IBM quantum computers. Before we get into the details of our attempts to observe quantum anomalous heat flow, it is useful to take a brief note of the relevant features of these NISQ devices. We focus on a class of devices that offer five to seven physical qubits, implementing different versions of the Falcon processor architecture \cite{IBMFalcon}. The IBM Falcon processors are a special class of quantum processors that, through a series of revisions, were eventually declared stable to mid-circuit measurements in early 2021. Moreover, the improved Falcon architecture also forms the basis for the more recent Eagle quantum processors which support up to 127 qubit devices \cite{IBMEagle}. We carried out our heat flow experiments on several IBM quantum computers based on the Falcon architecture, but for this paper, we focus on the results obtained from three representative devices listed in Table~\ref{tab:IBMQcomps}. 
\begin{table}
\centering
    \begin{tabular}{|c|c|c|c|}
    \hline
    Device & No. of & Processor & Optimized for \\
    name & qubits &  & mid-circuit \\
    & & & measurements \\
    \hline
    \emph{ibmq$\_$quito} & 5 & Falcon r4T & No \\
    \emph{ibmq$\_$manila} & 5 & Falcon r5.11L & Yes \\
    \emph{ibm$\_$perth} & 7 & Falcon r5.11H & Yes \\
    \hline
    \end{tabular}
\caption{IBM quantum computers used in this work}
\label{tab:IBMQcomps}
\end{table}

\section{Experimental Results}
\label{sec:results}
To highlight the importance of a quantum computer's stability under mid-circuit measurements for observing quantum anomalous heat flow, we begin by discussing the data obtained from \emph{ibmq$\_$quito}. The quantum processor \emph{ibmq$\_$quito} has a total of five qubits managed by an earlier version of the Falcon processor (Falcon r4T) which was not optimized to handle mid-circuit measurements. An attempt to observe the quantum anomalous heat flow
on this device leads to the results presented in Fig.~\ref{fig:exp_quito}. In this figure, the blue curves show the theoretical estimates (same as Fig.~\ref{fig:th}) while the red one shows the raw data obtained from \emph{ibmq$\_$quito}. Fig.~\ref{fig:exp_quito}(a) and (b), respectively, show the quantum heat flow ($Q_{\rm q}^c/\Delta$) and the semi-classical heat flow measured within the two-point-measurement scheme ($Q_{\rm sc}^c/\Delta$) against the evolution parameter $\theta$.

While the experimental data in Fig.~\ref{fig:exp_quito} generally does not agree quantitatively with the corresponding theoretical predictions, the difference between the quality of the experimental data for ($Q_{\rm q}^c/\Delta$) and ($Q_{\rm sc}^c/\Delta$) is conspicuous and can safely be attributed to the use of mid-circuit measurements in the determination of $Q_{\rm sc}^c/\Delta$ (see Fig.~\ref{fig:qct_tpm_circuit}). In particular, the noise in $Q_{\rm sc}^c/\Delta$ makes the observation of quantum anomalous heat flow on \emph{imbq$\_$quito} inconclusive (see Fig.~\ref{fig:exp_quito}(c)). In an attempt to improve these results, we used a mitigation scheme that corrects for the statistical error in detecting the state of a qubit (detecting a $|0\rangle$ when it is actually $|1\rangle$ and vice versa) while assuming that such errors are not affected by inter-qubit interactions. This type of error is often referred to as state preparation and measurement (SPAM) error and can be efficiently mitigated using a separate set of calibration circuits \cite{bravyi2021mitigating}. The data thus mitigated is shown in green in Fig.~\ref{fig:exp_quito}. Clearly, this simple error mitigation scheme hardly improves the quantitative and qualitative differences between the theory and the observation.

\subsection{Experimental detection of the anomalous quantum heat flow}
\begin{figure}
    \centering
\includegraphics[width=1.0\columnwidth]{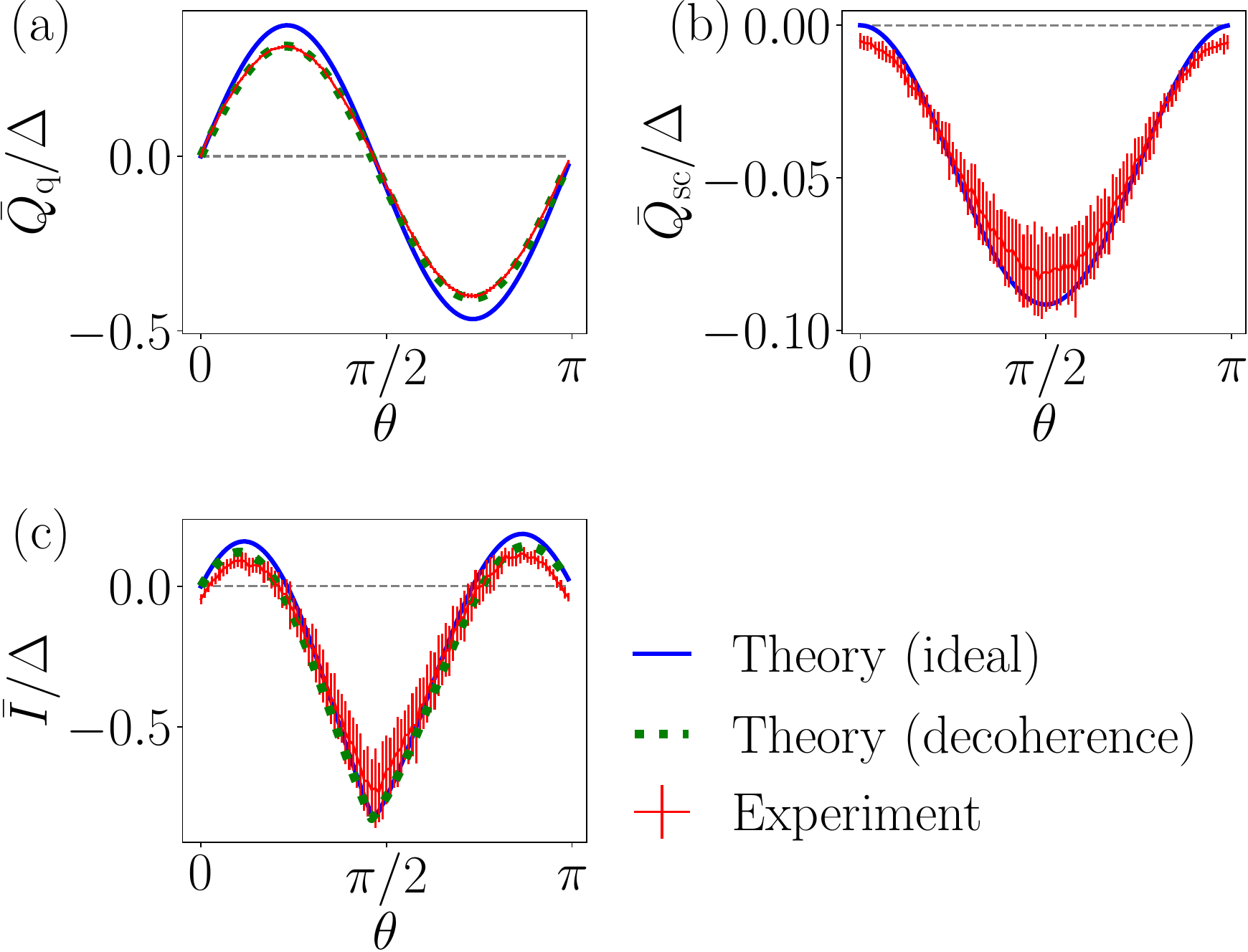}
    \caption{{\it Experimental results from ibmq$\_$manila averaged over 15 runs}: For $\beta_c \Delta =0.51$ and $\beta_h \Delta =0.13$, the red traces in (a) and (b) show the average $\bar{Q}_{\rm q}/\Delta$ and $\bar{Q}_{\rm sc}/\Delta$, respectively, as a function of $\theta$. The red bars are the standard deviations in the data. (c) shows the violation of the inequality in Eq.~(\ref{eq:2quibit_ineq}). The blue traces show the theoretical predictions. The green dotted lines are obtained when the decoherence parameter $\zeta$ is set to 0.06 (see text for further details).}
    \label{fig:avg_exp_manila}
\end{figure}

Let us now move on to an IBM quantum computer with improved capability to handle mid-circuit measurements, \emph{ibmq$\_$manila}. This five-qubit device is managed by the Falcon r5.11L processor, which was optimized to support mid-circuit measurements (see Table~\ref{tab:IBMQcomps}). The `L' in Falcon r5.11L refers to the linear layout of the qubits, which is optimal for the present implementation \footnote{This choice seems to be critical as we were not able to observe a quantum anomalous flow on {\it ibm}$\_${\it perth} processor, which belongs to the class of Falcon r5.11H and has qubits in an ``H'' layout.}. Results obtained from this device are presented in Fig.~\ref{fig:avg_exp_manila}, where the data is averaged over 15 realizations and the error bars represent the standard deviation. In Fig.~\ref{fig:avg_exp_manila}(a) we compare the theoretical (blue curve) and experimental (red curve) values of the mean quantum heat flow,
\begin{equation}
    \bar{Q}_{\rm q} \equiv (Q_{\rm q}^c - Q_{\rm q}^h)/2,
    \label{eq:Qq-bar}
\end{equation}
at different values of $\theta$. Here, the theory and the experiment are in good qualitative agreement with each other. The small quantitative difference is addressed below.

\begin{figure}
    \centering
\includegraphics[width=1.0\columnwidth]{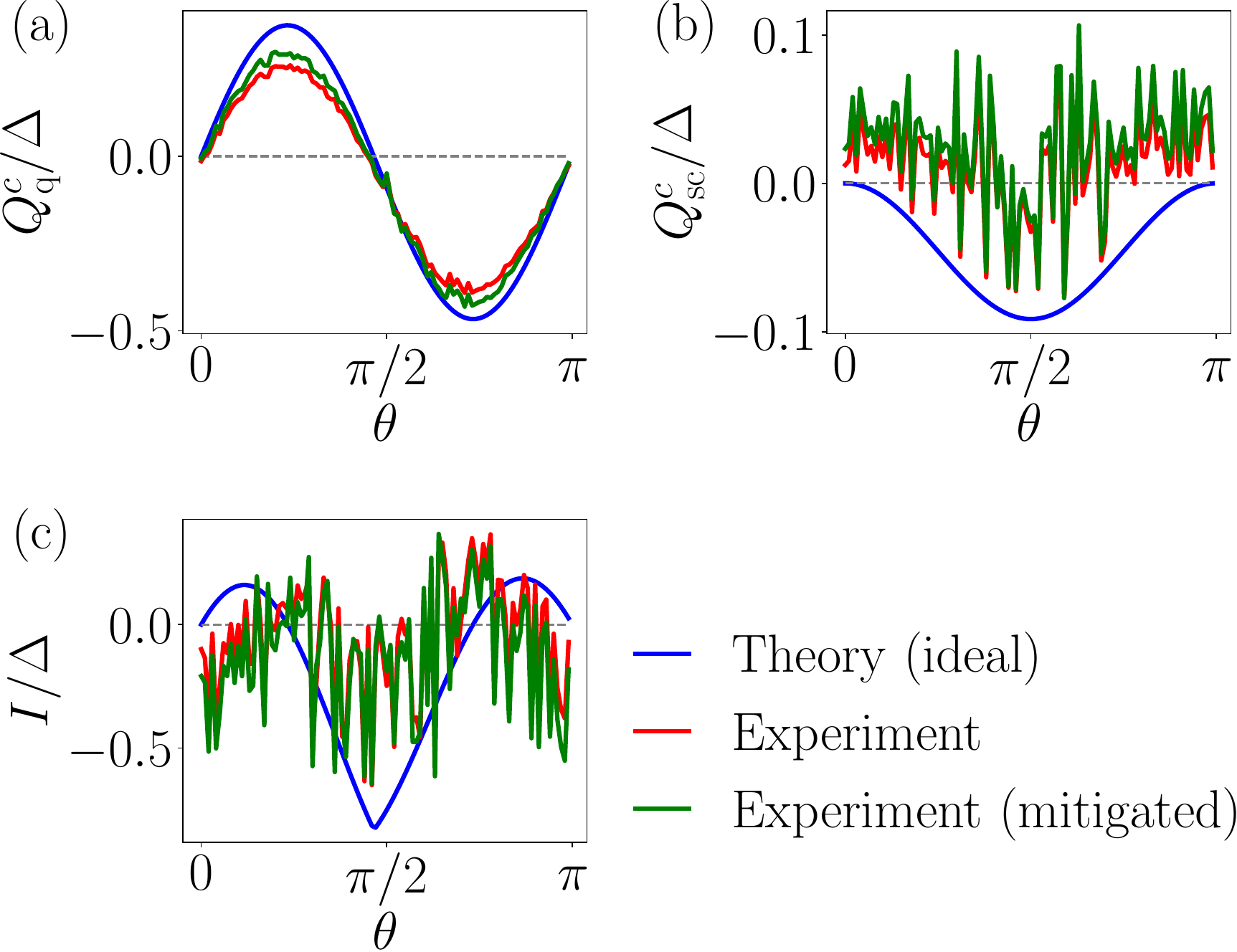}
    \caption{{\it Experimental results from ibmq$\_$quito}: For $\beta_c \Delta =0.51$ and $\beta_h \Delta =0.13$, (a) and (b) show $Q_{\rm q}^c/\Delta$ and $Q_{\rm sc}^c/\Delta$, respectively, as a function of $\theta$. (c) shows the violation of the inequality in Eq.~(\ref{eq:2quibit_ineq}). In these plots, the red and the green traces, respectively, show the raw data and the corresponding error-mitigated data (details in the text). While the blue traces show the theoretical predictions. The effect of mid-circuit measurements in (b) and (c) is clearly quite strong.}
    \label{fig:exp_quito}
\end{figure}

In Fig.~\ref{fig:avg_exp_manila}(b) we show the mean semi-classical heat flow
\begin{equation}
    \bar{Q}_{\rm sc} \equiv (Q_{\rm sc}^c - Q_{\rm sc}^h)/2
    \label{eq:Qsc-bar}
\end{equation}
within the two-point measurement scheme as a function of $\theta$. As before, the blue curve shows the theoretical prediction while the red curve is the experimental observation. In clear contrast to the corresponding observation from \emph{ibmq$\_$quito} (Fig.~\ref{fig:exp_quito}(b)), here we find that the noise is significantly reduced and the agreement with the theoretical predictions is also reasonably good. Finally, in Fig.~\ref{fig:avg_exp_manila}(c) we plot the mean quantum violation $\bar{I}$ obtained by replacing $Q_{\rm q}$ and $Q_{\rm sc}$ in Eq.~(\ref{eq:2quibit_ineq}) by their mean values $\bar{Q}_{\rm q}$ and $\bar{Q}_{\rm sc}$, respectively. Owing to the stability of \emph{ibmq$\_$manila} to mid-circuit measurements, one is now able to make a clear observation of the quantum anomalous heat flow.

It is worth noting that Fig.~\ref{fig:avg_exp_manila}(a) is similar to Fig.~\ref{fig:exp_quito}(a) which was obtained from \emph{ibmq$\_$quito}. In fact, the slight quantitative difference between the theory and the experiment in these plots is a generic feature present in the data for $Q_{\rm q}/\Delta$ obtained from all the IBM quantum computers 
where we performed our heat flow experiments and not just the representative ones listed in Table~\ref{tab:IBMQcomps}. This generic feature can be modeled by adding a single decoherence parameter $\zeta$, which is independent of $\theta$, to the coherence term $\eta$ in the density matrix of the initial state $\rho_{ch}$ (see Eq.~(\ref{eq:init_state})). In Fig.~\ref{fig:avg_exp_manila}, we use the dotted green curves to show the theoretical prediction for $\zeta = 0.06$ and find them to be in good agreement with the experimental (red) curves. Noting that our choice of the initialization parameters, $t_c = \pi/3$ and $t_h = \pi/6$ gives $\eta \simeq -0.42$ (see Eq.~\ref{eq:eta}), a small positive $\zeta$ can indeed represent the loss of coherence in the initial state. Furthermore, since $\zeta$ only contributes to an off-diagonal term of the initial density matrix, it drops out in the two-point measurement scheme and thus, does not affect the theoretical prediction for the semi-classical heat flow (see Fig.~\ref{fig:avg_exp_manila}(b)).

Our experimental observations on \emph{ibmq$\_$manila} suggest that the Falcon r5.11 quantum processor and its successors are indeed stable to mid-circuit measurements. However, we have also uncovered a couple of issues which object to this conclusion. These are what we describe next.

\begin{figure}
    \centering
\includegraphics[width=1.0\columnwidth]{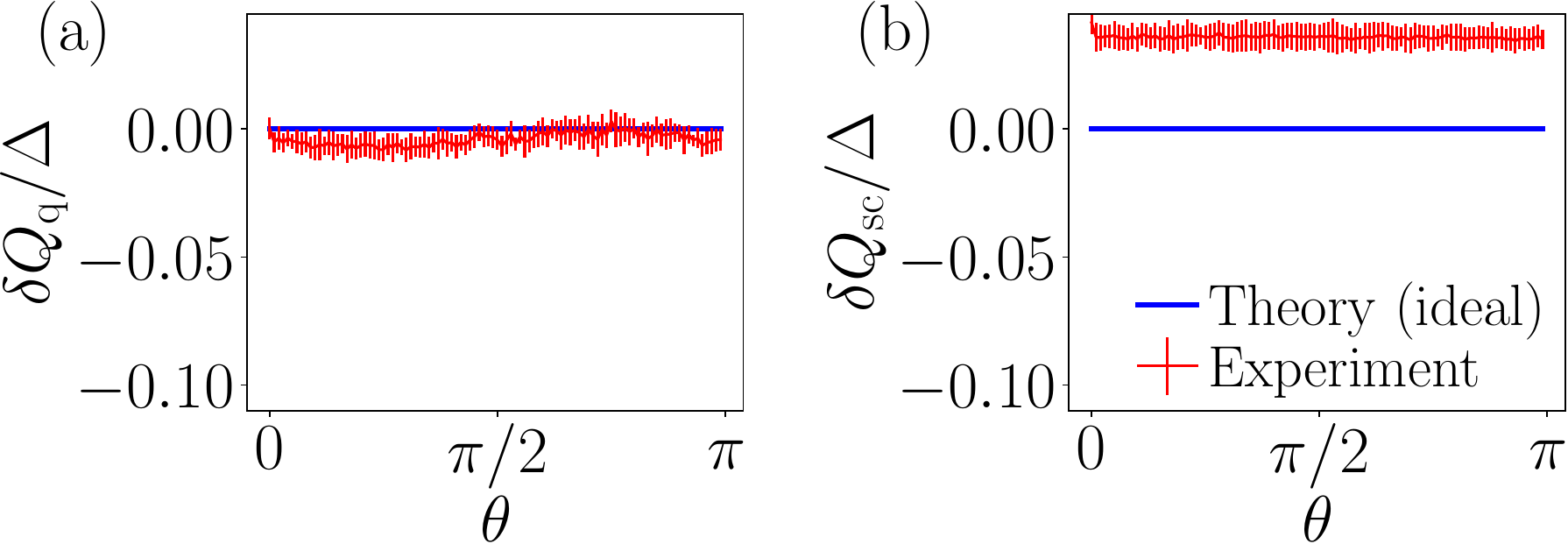}
    \caption{{\it Experimental observation of deviations from the expected energy conservation (data from ibmq$\_$manila averaged over 15 runs)}: For $\beta_c \Delta =0.51$ and $\beta_h \Delta =0.13$, the red traces in (a) and (b) show the average $\delta Q_{\rm q}/\Delta$ and $\delta Q_{\rm sc}/\Delta$, respectively, as a function of $\theta$. The red bars are the standard deviations in the data. The blue traces in (a) and (b), respectively, show the theoretical predictions for $\delta Q_{\rm q}/\Delta$ and $\delta Q_{\rm sc}/\Delta$.}
\label{fig:manila_discrepency}
\end{figure}

\subsection{Effect of mid-circuit measurements on total energy conservation}
To further benchmark the effect of mid-circuit measurement, we now probe the variations in the total energy of the system,
\begin{equation}
    \delta Q_{\rm q(sc)} \equiv (Q_{\rm q(sc)}^c + Q_{\rm q(sc)}^h) \ .
    \label{eq:dE}
\end{equation}
Ideally, both $\delta Q_{\rm q}$ and $\delta Q_{\rm sc}$ are expected to be zero, because the unitary operator $U_\theta$ conserves the total energy of the two qubits $c$ and $h$, Eq.~(\ref{eq:energy_cond}). In Fig.~\ref{fig:manila_discrepency}(a) and (b), respectively, we present the experimental observations for $\delta Q_{\rm q}/\Delta$ and $\delta Q_{\rm sc}/\Delta$ averaged over fifteen independent runs on $ibmq\_manila$. 
We find that both $\delta Q_{\rm q}$ and $\delta Q_{\rm sc}$ are essentially independent on $\theta$.
Moreover, 
$\delta Q_{\rm q}$ agrees with the theoretical prediction closely implying that there is no significant energy flow in or out of the system during the measurement of $Q_{\rm q}$. This also implies that all the unitary operations involved in the experiments are largely functioning as expected.
In contrast, $\delta Q_{\rm sc}/\Delta$ in Fig.~\ref{fig:manila_discrepency}(b) deviates considerably from the theoretical prediction implying a significant amount of energy flowing out of the system during the measurement of the semi-classical heat flow $Q_{\rm sc}$. The pronounced difference between the behavior of $\delta Q_{\rm q}$ and $\delta Q_{\rm sc}$
strongly suggest that the source of this energy loss from the system is the imperfect implementation of the additional process employed in the observation of $Q_{\rm sc}/\Delta$, namely, mid-circuit measurements.

\begin{figure}
    \centering
\includegraphics[width=1.0\columnwidth]{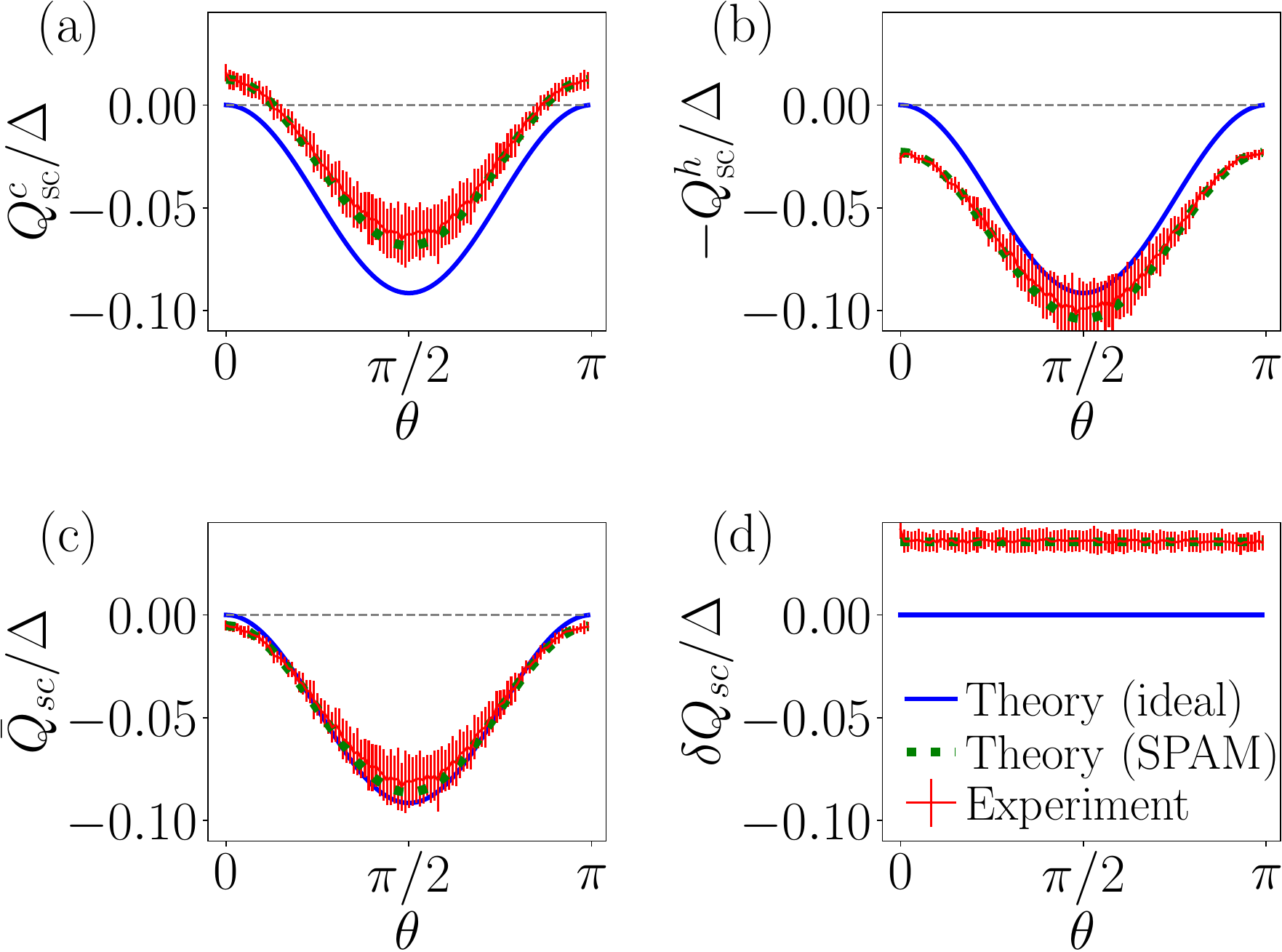}
    \caption{{\it Our model fit to $Q_{\rm sc}$ from ibmq$\_$manila averaged over 15 runs}: For $\beta_c \Delta =0.51$ and $\beta_h \Delta =0.13$, (a), (b), (c) and (d) show $Q_{\rm sc}^c$, $-Q_{\rm sc}^h$, $\bar{Q}_{\rm sc}$, and $\delta Q_{\rm sc}$, respectively. The blue curves show the (ideal) theoretical predictions while the green dotted lines show the results from our model incorporating the effect of mid-circuit-measurements (see text for details). The red curves show the experimental values averaged over 15 runs while the red bars show the standard deviation in the data. Our model is in good agreement with the experimental observations.}
    \label{fig:model_fit}
\end{figure}

The main source of disturbance in the measurement process of superconducting quantum circuits is related to a systematic bias towards the detection of the $|0\rangle$ state. We can model this type of SPAM error by adding and subtracting a fixed amount $\delta_{c(h)}$ to the diagonal terms of the density matrix, leading to the new density matrix
\begin{equation}
    \tilde{\rho}_{c(h)} = \rho_{c(h)} + \delta_{c(h)} \sigma_z
    \label{eq:rho_c_tilde}
\end{equation}
where $\sigma_z = {\rm Diag}\{1,-1 \}$ is a Pauli matrix and $\delta_{c(h)}$ is a real number.
We next note that although $\mathcal{D}\rho_{ch} \neq \rho_{c} \otimes \rho_{h} $ ($\mathcal{D}$ is the two-measurement decoherence operator defined after Eq.~(\ref{eq:Qsc})), one can use the latter to measure the semi-classical heat flow: As shown in Ref.~\cite{2020_PRXQ_Levy},
\begin{equation}
Q^{c(h)}_{\rm sc} = {\rm Tr} \left[ \rho_{ch} H_{c(h)} - U_\theta \rho_{c} \otimes \rho_{h} U^\dagger _\theta H_{c(h)}\right]\ .
\label{eq:formula_Qsc_new}
\end{equation}
By using $\tilde{\rho}_{c(h)}$ instead of $\rho_{c(h)}$ in Eq.~(\ref{eq:formula_Qsc_new}) we find that
\begin{align}\delta Q_{\rm sc}/\Delta = \delta_c + \delta_h\end{align}
is non-zero and independent of the evolution parameter $\theta$. This result is in good agreement with the experimental measurement of $\delta Q_{\rm sc}/\Delta$ presented in Fig.~\ref{fig:manila_discrepency}(b). In fact, as shown in Fig.~\ref{fig:model_fit}, for $\delta_c = 0.0125$ and $\delta_h = 0.023$ one can capture the average behavior of both $Q_{\rm sc}^{c}/\Delta$ and $Q_{\rm sc}^{h}/\Delta$, and consequently of $\bar{Q}_{\rm sc}/\Delta$ and $\delta Q_{\rm sc}/\Delta$, with considerable success.

\section{Summary and discussion}
\label{sec: summary}

This study presents the first observation of quantum anomalous heat flow on quantum computers. We utilized mid-circuit measurements to design quantum circuits that exhibit a violation of the heat flow inequality. This violation witnesses negativities in the real part of the Kirkwood-Dirac quasiprobability distribution, thereby identifying unique quantum phenomena during the process. These anomalous heat flows are attributed to the initial quantum correlations between the subsystems. 

Mid-circuit measurements, a feature that is now available on some quantum computing platforms, enable the probing of selected qubits during experiments, ideally without disturbance of the quantum state. This capability is crucial for measuring heat flow using a two-point measurement scheme, which is essential for assessing the heat-flow inequality.
However, we find that the primary discrepancies between the experimental results and the theoretical predictions are actually due to mid-circuit measurements. These measurements introduce disturbances that result in energy leakage and biased outcomes as was observed by the total heat flow variation. We modeled this discrepancy in the measurement process by adjusting the population distribution.

Mid-circuit measurements are expected to be critical for the advancement of quantum thermodynamics on NISQ devices and for broader quantum computing applications. However, the precise characterization of errors introduced by these measurements remains under study. Continued improvements in this area are necessary to enhance the reliability and performance of quantum technologies. Our work demonstrates the use of quantum thermodynamics to characterize the limitations of state-of-the-art quantum computers, offering a promising route to benchmark and improve these devices.

\subsection*{Ackowledgment} This work was supported by the Israel Science Foundation, grant numbers 151/19, 154/19, and 2126/24. 
A. Levy acknowledges support from the Israel Science Foundation, Grant No. 1364/21 and 3105/23.

\bibliography{biblio}
\end{document}